\documentclass[useAMS,usenatbib,usegraphicx]{mn2e}
\usepackage{epsf}

\title[Thermal emission in GRB afterglows]{Thermal emission in gamma-ray burst afterglows }
\author[D. A. Badjin, S. I. Blinnikov and K. A. Postnov]{
D. A. Badjin$^{1}$\thanks{E-mail:badjinda@gmail.com}, S. I.
Blinnikov$^{2,1}$
\thanks{E-mail:sergei.blinnikov@itep.ru}
and K. A. Postnov$^{1}$
\thanks{E-mail:kpostnov@gmail.com}
\\
$^{1}$Moscow  M.V. Lomonosov  State University, Sternberg Astronomical Institute, Universitetsky prosp., 13, Moscow, 119992, Russia\\
$^{2}$Institute for Theoretical and Experimental Physics, 117259,
Moscow, Russia}
\begin{document}

\date{Accepted ...; Received ...; in original form 22 January 2013}

\pagerange{\pageref{firstpage}--\pageref{lastpage}} \pubyear{2013}

\maketitle

\label{firstpage}

\begin{abstract}

We study thermal emission from circumstellar structures heated by gamma-ray burst (GRB) radiation and ejecta and calculate its contribution to GRB optical
and X-ray afterglows using the modified radiation hydro-code {\small STELLA}.
It is shown that thermal emission originating in heated dense shells around the GRB progenitor star can reproduce X-ray plateaus (like observed in GRB 050904, 070110)
as well as deviations from a power law fading observed in optical afterglows of some GRBs (e.g. 020124, 030328, 030429X, 050904). 
Thermal radiation pressure in the heated circumburst shell dominates the gas pressure, producing rapid expansion of matter similar to supenova-like explosions   
close to opacity or radiation flux density jumps in the circumburst medium. This phenomenon can be responsible for so-called supernova bumps in optical afterglows
of several GRBs. Such a `quasi-supernova' suggests interpretation of the GRB-SN connection which does not directly involve the explosion of the GRB progenitor star.

\end{abstract}


\section{Introduction}

Cosmological gamma-ray bursts (GRBs) are the most luminous events in the Universe, which are observed as short flashes of gamma-ray radiation (prompt emission)
accompanied by transient afterglow on longer wavelengths (see \citealt*{GRF} for a review and references).

After decades of intensive studies, it has been commonly accepted that GRB radiation has mainly a non-thermal nature. The principal prompt emission mechanisms include 
synchrotron radiation of electrons accelerated in relativistic shock waves or magnetohydrodynamic (MHD) instabilities, synchrotron self-Compton and inverse Compton scattering of thermal photons 
(see e.g. \citealt{Beloborodov10,Daigne11,Zhang}). 

However, thermal emission can also emerge in GRB light curves for several reasons.
It is commonly accepted that long GRBs 
are connected with explosive deaths of massive stars, which can be 
surrounded by a dense stellar wind or even by shells of matter 
(up to several $M_{\odot}$) produced by pulsational instability of the progenitor. 
Even several per cent of the huge GRB energy $\sim 10^{50}-10^{53}$~erg
intercepted by this matter
can be observed as thermal emission \citep{BP98}.

Thermal emission signatures have been found in GRB spectra. 
\citet{RydePeer} calculated a photospheric component 
on top of the non-thermal power-law, which can describe several tens of observed GRB prompt emission spectra better than a pure broken power law \citep{B93} or a cut-off power law, 
i.e. the `canonical' spectral functions.
Several detections of X-ray lines \citep*{Anton, Reeves02} and blackbody components (see e.g. \citealt*{Campana06}, \citealt*{Page11}) were reported in X-ray afterglows suggesting 
that thermal effects may also contribute to the afterglow radiation. 

\citet*{BKP} discussed how the power-law GRB spectra may be produced by the blackbody radiation with changing temperature. 
Recently, studies by several authors (e.g. \citealt*{Lazzati09}, \citealt*{Miz11}, \citealt{Nagakura11, Suz13}) make the GRB thermal prompt emission more topical. 
Using axisymmetric relativistic hydrodynamical simulations, they argue that, breaking through the progenitor star 
envelope, the relativistic jet becomes collimated and cut apart by tangential shock waves. Different parts of the jet start moving with different Lorentz-factors (an 
order of magnitude exceeding those predicted by the spherically symmetric fireball model) and, when colliding with each other, they can produce thermal emission. This looks 
similar to the widely accepted internal shock scenario \citep{Rees94, Piran99}, but the emerging thermal emission ensures greater efficiency than the synchrotron one. 

In addition, late-time afterglow light curves of GRBs 
often demonstrate bumps with colour evolution and optical and NIR spectra 
looking like type Ib/c supernova on top of the power-law fading.
These supernova-like features have been detected photometrically and spectroscopically in many GRB afterglows (see \citealt{WB06}, \citealt{Cano11} for a review of GRB-SN 
physical and observational connections), and are considered as the most compelling evidence for the presence of thermal plasma radiation in GRB afterglows. This GRB-SN 
connection is a cornerstone of the present concepts of the origin of GRB central engine.

Therefore, thermal emission in GRB afterglows can appear as irregular, `sporadic' deviations from the `canonical' power laws of GRB light curves
and spectra $F_\nu\sim t^{-\alpha}\nu^{-\beta}$.

Thermal emission from large circumburst structures has also been studied. For example,
\citet{BKT, BBK} studied the case where a dense molecular cloud around and near the GRB progenitor star was present. In their simulations, 
a cloud with number density $n\sim 10^5$~cm$^{-3}$ and radius of a few parsec was heated up by an instant (delta-like) gamma-ray emission pulse via Compton 
scattering, and the generated thermal radiation was found to last for a few years.

In our previous paper (\citealt*{BBP}, hereafter Paper I), we modelled thermal effects
in GRB afterglows using the multigroup radiation-hydrocode {\small STELLA}. The code was 
originally developed for detailed simulations of supernova radiation (see \citealt{B98}), and is suitable for calculation of thermal emission generation and transport in 
exploding and expanding media. In paper I the code was modified to include gamma-ray heating and non-stationary ionization of matter. 

In the present paper, the code is modified in order to more accurately handle 
discontinuities (in degree of ionization, opacity and emissivity, etc.) 
which emerge due to the propagation of gamma-ray radiation. The GRB jet interaction with the shell is modelled using the so called `thermal bomb' 
(in contrast to a heavy `piston', or `quasi-ejecta' which we used in Paper I). In this approach a certain amount of thermal energy is  deposited into a certain 
amount of mass during a certain time interval. Different boundary conditions can be imposed to thermal radiation escaping from the shell.

In Section 2 we describe the model. Section 3 presents results of calculations.
In Section 4 we compare the results of our calculations with features of observed afterglows and discuss
other possible astrophysical applications. Section 5 summarizes the results obtained.

\section[]{The model}

The primary goal of our calculations is to study thermal emission generated during the interaction of 
powerful gamma-ray flux and following blast wave from a GRB with the cold dense shell surrounding the GRB progenitor and to find possible 
observational signatures of this radiation. 

A massive GRB progenitor star can experience a powerful pulsational instability stage before the explosion and expell 
several M$_{\odot}$ of matter. 
This scenario, for example, was suggested by \citet*{WBH} to explain an extremely bright Type IIn supernova SN2006gy. Observational signatures of 
such mass-loss events around supernova progenitors were recently found by \citet{Ofek}.

Like in \citet{WBH}, in our calculations we used the multigroup radiation hydrocode {\small STELLA}.
The code is one-dimensional and non-relativistic, but radiative transfer equations include aberration, time delay and Doppler-shift effects to within the $O(v/c)$ accuracy.  
Therefore, to aviod relativistic motions, we are restricted by spherically symmetric shells with densities corresponding a total Thomson optical depth of less than one 
and located far enough from the centre of explosion. 
Specifically, we consider thin  shells at a distance of $\sim10^{16}$~cm with number densities of several $10^{10}$~cm$^{-3}$. 

As \citet{WBH} show, such and even denser `walls' 
around the GRB progenitor could arise if the central star had experienced several mass ejection episodes, and the ejecta 
interacted with each other.

\subsection[]{Initial parameters}

Let us consider a spherical shell with mass 5 M$_{\odot}$, radius $R=10^{16}$~cm and thickness $5\times$10$^{13}$~cm. 
The mean baryon number density in the shell is $n_{\rmn{bar}}\sim 10^{11}$~cm$^{-3}$, which is high enough to produce considerable thermal effects and keep 
the dynamics non-relativistic. The relatively small shell thickness allows a good spatial resolution. The elemental abundance was assumed to be the same as in the outer 
shells of the pre-supernova calculated by \citet{WBH}. 

The shell is illuminated by gamma-ray radiation with a peak isotropic 
luminosity of 3$\times$10$^{53}$ erg s$^{-1}$ consisting 
of three FRED (Fast Rise -- Exponential Decay) 
pulses, each with the characteristic duration 1.3 s, 
so the total energy of gamma-ray emission was about 
4.5$\times$10$^{53}$ erg. These luminosity and the gamma-ray peak duration 
are quite typical (see, e.g.,  statistical study in \citealt*{BBG}). 
The gamma-ray light curve was intentionally made `spiky' to see the possible 
quick response of the surrounding shell to such rapidly altering
 `perturbations' (no effect was actually found).

The spectral energy distribution of the 
gamma-ray emission was taken in the form of a broken power law (the Band function, \citealt{B93}) 
with the characteristic energy $\epsilon_0=300$ keV and low- and high-energy exponents 0.9 
and 2, respectively. The half opening angle of the gamma-ray emission cone was 
taken to be $\theta_{\rmn{jet}}=10^\circ$. To treat the emergent thermal emission,
120 frequency groups from 50000{\AA} in IR to 30 keV in X-rays
were used.

\subsection[]{Abilities of the code}

For the present calculations several modifications were implemented to the {\small STELLA} code. 
To treat the interaction of gamma-rays with matter, a time-dependent system of ionization balance equations is solved. Photoionization, 
ionization due to Compton scattering on bound electrons, collisional ionization, radiative and dielectronic recombination are taken into account. These elementary processes
determine the effect of the gamma-ray emission on the state of matter, its opacity and emissivity. 

The opacity of the shell matter should not block the non-thermal synchrotron emission of relativistic shocks in the GRB ejecta (in order to observe the GRB phenomenon itself). This 
requirement is fulfilled because the full ionization of matter by gamma-rays strongly reduces its photoelectric opacity. A large value of the ionization parameter 
$\xi=978\,n_{10}^{-1}$ for photon energies above 10~keV and $1760\,n_{10}^{-1}$ for those above 1~keV ($n_{10}$ is the bound electron number density 
in units $10^{10}$~cm$^{-3}$) indicates that there is a lot of high energy  photons per every bound electron. Our numerical non-stationary ionization calculations 
show that the matter becomes fully ionized in the first thousandths of a second after the gamma-ray forward front. Only a few hydrogen-like ions of heavy elements 
(e.g. Fe\,{\small XXVI}) with the highest recombination rates remain in non-negligible amounts of $0.1-10$~cm$^{-3}$. Thus, both the prompt non-thermal GRB emission and 
the early X-ray afterglow will be visible until the recombination of matter occurs. 

The radiative heating of cold matter is produced by the gamma-ray radiation via photoionization and Compton scattering (with opacity coefficients depending on 
the non-stationarily calculated state). In addition to the heating effect, the radiation drag force produced by gamma-radiation is taken into account.

As the radiation front moves with the speed of light, the unheated regions are not causally connected to heated ones; there is a sharp discontinuity 
in temperature, ionization, opacity and emissivity behind the gamma-ray front. No outward fluxes of heat, mass and radiation  
through this contact discontinuity are assumed as boundary conditions. In the modified code this is taken into account.
When calculating the Eddington factors of the thermal emission (see \citealt{B98}), the heated and unheated parts of the shell are 
treated independently.

Due to high densities and radiation fluxes (both thermal and non-thermal), the radiation pressure can dominate the gas pressure by four orders of magnitude. 
This makes the problem 
very sensitive to the local opacity, direction and strength of radiation flows. Depending on whether there are highly opaque regions (where the radiation is strongly coupled 
with matter) or how transparent the inner boundary is (we mention here only the inner boundary, since the flux conditions on the outer and intermediate ones are set strictly 
as `outwards only' and 'inwards only'), the radiation either flows away exhausting shell's thermal energy, or is accumulated in hot regions and drives their expansion. 

The latter, for example, takes place if the inner boundary for some reason does not transmit the radiation towards the centre (e.g. it is highly compressed by a shock and 
rather opaque, or scatters the radiation back, or due to relativistic aberration makes most of the emission flow outwards). Then the thermal radiation appears trapped 
between the `no-flux-towards-the-centre' inner and 'no flux outwards' intermediate boundaries, accelerating the hot zones rapidly up to several $10^4$~km~s$^{-1}$, which
results in a `quasi-supernova' event (see Section 4). As matter is needed to be compressed by several orders of magnitude to compensate 
the radiation pressure, a radiation implosion easily develops around the opacity or radiation flux jumps. Note that the compression increases the opacity, 
providing a positive feedback on the implosion, until the radiative pressure growth coasts or some instabilities come into play smearing off the density peak.

We place the inner boundary flux condition by means of the inner boundary Eddington factors. 
They are calculated similarly to those at the outer boundary in the original {\small STELLA} code (${\mathcal H}_\nu=h_{\rmn{E}}{\mathcal J}_\nu$ in terms of \citealt{B98}).
An arbitrary incoming boundary flux, e.g. a diluted blackbody background or a counter-jet emission, can also be taken into account.


In addition to being heated by gamma-rays, the shell is heated due to partial thermalization of the GRB jet kinetic energy. 
Clearly, the one-dimensional non-relativistic code cannot model 
the relativistic jet dynamics self-consistently; therefore we model the jet impact by means of a `thermal bomb'.
In this approach, which is widely used in simulations of supernovae, the jet braking and its kinetic energy dissipation are described 
in terms of the thermal energy which is deposited into selected zones during the time interval mostly appropriate for the effecive 
dissipation process.

This general approach 
is rather insensitive to details of the GRB fireball expansion or blast wave propagation.
Assuming energy equipartition between the gamma-rays and ejecta, we have taken $4.5\times10^{53}$~erg as an estimate of the energy to be thermalized. 
In fact, results of calculations turned out to be sensitive to the total thermalized energy and weakly depend on deviations from the equipartition. 

In order to determine where to deposit the energy in the shell, the shock Lorentz factor should be known. The `thermal bomb' domain should
have mass exceeding $E_{\rmn{sh}}(c\Gamma_{\rmn{sh}})^{-2}$ [the characteristic swept mass scale at which the relativistic self-similar regime \citep{BMK} is established], 
but less than $E_{\rmn{sh}}c^{-2}$ (the characteristic scale of the non-relativistic Sedov-Taylor regime). The blast wave moving with the Lorentz-factor $\Gamma_{sh}$
impacts the shell inner boundary at the time $\delta t_{\gamma-\rmn{sh}}=R_{\rmn{shell}}(2c\Gamma^2_{\rmn{sh}})^{-1}$ after gamma-rays. 
The duration of the heating process can be estimated as the radial thickness of zones in which the energy is deposited divided by the speed of light.

We assumed the blast wave Lorentz-factor $\Gamma_{\rmn{sh}}\approx30$. This is suitable for the incoming shock with the initial fireball Lorentz-factor
$\Gamma_{\rmn{sh,i}}\sim10^{2.5-3}$ which follows the Blandford-McKee solution with $\Gamma_{\rmn{sh}}\propto R^{-3/2}$ in a homogeneous medium for an adiabatic shock (or 
$\Gamma_{\rmn{sh}}\propto R^{-3}$ for a radiative shock) well before the impact. This means that several 
$E_{\rmn{sh}}(c\Gamma_{\rmn{sh,i}})^{-2}\approx2.5\times10^{-5}-10^{-6}\rmn{M}_{\odot}$ and $\ll E_{\rmn{sh}}c^{-2}\approx0.25\rmn{M}_\odot$ (in order for 
the shock to be relativistic)
of circumburst matter should be swept up by that moment. This indeed could be the case if the shell itself were produced by a non-relativistic shock initiated by one 
non-relativistic ejecta moving through another (like in a pair-instability supernova, \citealt{WBH}).

Thus, with $\Gamma_{\rmn{sh}}\approx30$ we have placed the `thermal bomb' within $20E_{\rmn{sh}}(c\Gamma_{\rmn{sh}})^{-2}\approx0.05\rmn{M}_\odot$ of shell's matter from its 
inner boundary. This also corresponds to the first Lagrangean zone. The `thermal bomb' is `ignited' $\delta t_{\gamma-\rmn{sh}}\approx200$ seconds after the radiative
heating begins in the first zone and is active for 17 s (which also coincides with the prompt emission duration). 
Hence, there are two kinds of heating processes involved: the radiative heating which has 
the 17~s transient profile and moves outwards from zone to zone with the speed of light, and the kinetic one which has a 17-s rectangular shape delayed by 200~s 
and acts in the innermost zone of the shell.

The shell mass exceeds $E_{\rmn{sh}}c^{-2}$ significantly (by a factor of 20 in our case). If the shell had uniform density 
distribution per unit solid angle, the shock would become non-relativistic well inside it, and there would be no `canonical' long-term (several hours or days in observer's time-scale)
power-law synchrotron afterglow. Thus, to produce such an afterglow, the shell should be inhomogeneous, consisting of `clumps' (or filaments) which intercept energy, 
and rarefied `windows', through which the relativistic fluid could flow without deceleration. As the shell itself is assumed to be created by a non-relativistic shock
in the circumstellar medium, such `clumps' and `windows' may naturally form due to instabilities. Dense clumpy circumburst matter was also introduced by \citet{PBKS} to explain 
emission lines in the X-ray afterglow of GRB 011211 \citep{Reeves02}.

Note, however, that for an opaque (or reflecting) inner boundary, during several days
thermal emission can mimic the power-law decay $F_\nu\propto t^{-\alpha}$ with exponential power typical for observed GRB afterglows 
 $\alpha\sim1-2$ (see Section 4). 


Finally, some reduction procedures were added to take into account the jet beaming factor and the spherical shape of the shell: (1) the hotspot fraction is  
$(1-\cos\theta_{\rmn{jet}})/2$ of the initially spherical shell; (2) the geometric time delay of the emission is 
$$\delta t(t)=(1-\bmath{e_r\cdot n}_{\rmn{obs}})R_{\rmn{out}}\bigl(t-\delta t(t)\bigr)/c$$ 
where $R_{\rmn{out}}(t)$ is the radius of the outermost zone in which the thermal emission is calculated, 
$\bmath{e_r}$ and $\bmath{n}_{\rmn{obs}}$ are local unit vectors directed from the emitting point radially and along the line of sight, respectively;
(3) the angular distribution of the outcoming thermal emission (despite the main code is one-dimensional, an approximate angular distribution of radiation is repetitevily computed
to update Eddington factors).

The maximal geometric curvature delay is 
$\delta t =(1-\mu_{\rmn{jet}})R_{\rmn{out}}(t-\delta t)/c
\approx 5000 \rmn{s}\,,$ 
where $\mu_{\rmn{jet}}=\cos\theta_{\rmn{jet}}$. 
The jet axis can be slightly misaligned with the line of sight, 
leading to a more complicated time averaging. Thus, for example, in Paper I a $3^\circ$ off-set between these directions
was assumed, resulting in the smearing off the curvature delay over 3000--8000~s interval, 
but in the present calculations no misalignment is assumed to see thermal features more clearly.

On the illuminated spherical surface, 
a distant observer will see the emitting region 
as a set of concentric rings, 
each corresponding to the delayed surface luminosity.
At a particular moment, 
the observed isotropic luminosity equivalent 
will be the thermal luminosity in observer's direction 
averaged over the visible area with account of the time delay:
$$
L_{\nu,\rmn{iso}}(t)=8\upi^2 \int\limits_{\mu_{\rmn{jet}}}^1 I_{\nu}\bigl(\mu,t-\delta t(t,\mu)\bigr) R_{\rmn{out}}^2\bigl(t-\delta t(t,\mu)\bigr)\mu d\mu.
$$ 

To calculate the light curves seen by the terrestrial observer, a cosmological model with 
$H_0=73.5$ km s$^{-1}$ Mpc$^{-1}$, $\Omega_\Lambda=0.76$, $\Omega_{\rmn{m}}=0.24$ was used.

Several simplifying assumptions should be mentioned. First, a single-temperature fluid model is adopted for computational reasons. However, in a high-density medium 
(several $10^{10}$~cm$^{-3}$), the full equilibration establishes in the first tens of seconds (the corresponding $n_{\rmn{e}}t$ parameter reaches $10^{12}$~cm$^{-3}$s, 
\citealt{Vink11}). As the characteristic time-scales of the calculated light curves are significantly longer (thousands of seconds, see section 3), 
the one-temperature approximation is justified. 

Second, we neglect multiple scatterings of gamma-ray photons. This allows us to recalculate the radiation field at each time step without considerable increasing of the 
computation time (see Paper I for discussion). Therefore, the total absorbed gamma-ray energy can be somewhat underestimated. For a given gamma-ray spectrum, the  
total heating power due to the second scattering is nearly four times smaller than due to the first one. Since the shell is semi-transparent for Thomson scattering, 
the account for multiple-scattered photons can not increase the heating power significantly. However, it could affect outcoming gamma- and X-ray radiation spectra.

\section{Results of calculations}

Fig.~1 presents some physical characteristics of the shell matter during its heating: the gas temperature, its velocity and the thermal emission temperature 
$T_{\rmn{rad}}=\left(\frac{u_{\rmn{rad}}c}{4\sigma_B}\right)^{0.25}$, where $u_{\rmn{rad}}$ is the total energy density of the thermal emission. 
The profiles are shown for the time 800~s after the gamma-rays strike the shell.

Fig.~2 demonstrates the light curves of the thermal emission. Shown are the bolometric luminosity, 
X-ray (2-30 keV), soft X-ray (0.1-2 keV) and \textit{UBVRI} optical emission. 
For comparison, the characteristic range of the 
observed \textit{R}-band power-law afterglow light curves
\footnote{This range represents the upper and lower
boundaries of the observed GRB afterglows, which we constructed using about 4000 
data points compiled from the literature \citep{BBG}. 
See also Fig.~7 in \citealt{GRF} for illustration.} 
(corrected for the Galactic extinction and redshift) is also shown.
Spectral density profiles for several characteristic moments are presented in Fig.~3 (e.g. the initial state, peaks of emission in different ranges;
see the legend in Fig.~3). \textit{R}-magnitudes calculated for 
different redshifts (without extinction) are presented in Fig.~4. 
On these plots, time is counted from the moment at which the prompt emission front 
reaches the shell outer boundary, i.e. from the first pulse of the observed GRB. 

 \begin{figure}
 \includegraphics[width=84mm]{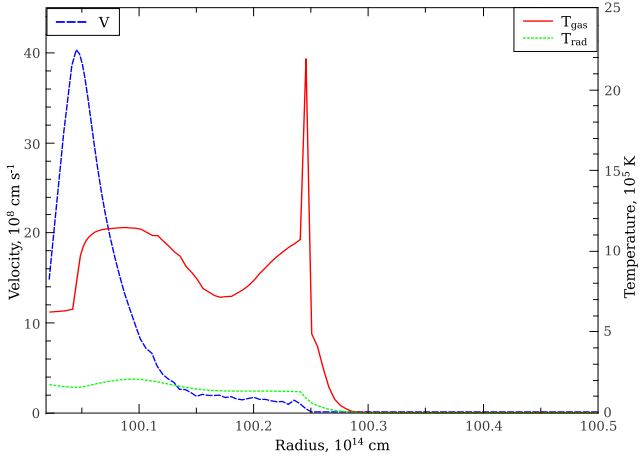}
 \caption{Radial profiles of the fluid velocity $V$, temperature $T_{\rmn{gas}}$ and the thermal emission temperature $T_{\rmn{rad}}$, calculated from its energy density,
    800~s after the heating started.}
\end{figure}

{\bf Gamma-ray effect}. After the sharp front of the gamma-radiation 
with isotropic peak luminosity $3\times10^{53}$ erg s$^{-1}$ 
impinges  on the cold (3000 K) dense shell, the following sequence of events occurs.
A nearly full ionization is reached in a few thousandths of a second since the photoionization 
rate dominates all other processes by several orders of magnitude. 
Only hydrogen-like ions of the heaviest elements are still present 
(0.1-10 ions per cm$^3$, while the total particle number density is of order $10^{11}$) due to recombination. 
This is important for the heat balance of matter and its thermal emissivity. 

\begin{figure}
 \includegraphics[width=84mm]{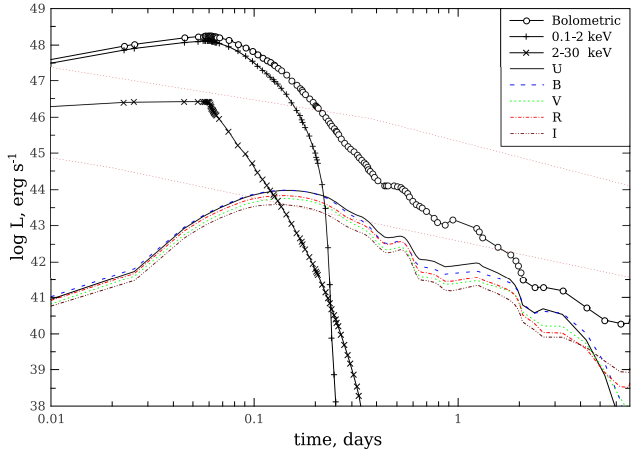}

 \caption{Isotropic equivalent luminosities 
of the calculated thermal emission. Bolometric (50000{\AA}--30 keV, open circles), 
X-ray (2-30 keV, inclined crosses), soft X-ray (0.1-2 keV, upright crosses), UBVRI (solid, dashed, dotted, dash-dotted and dash-dot-dot lines, respectively) light curves are shown. 
For comparison, the range of the observed 
\textit{R}-band power-law afterglows is plotted by the thin dotted straight lines.}
\end{figure}

As the material gets nearly fully ionized, the  
heating is mainly due to Compton scattering of gamma-rays off free electrons. 
The cooling is due to bremsstrahlung, inverse Compton scattering and radiative recombination which form the X-ray continuum.   

In 1~s after the radiative heating begins in a given fluid element, 
the dynamic balance of the medium and the gamma-radiation is reached 
at a temperature of  $\sim10^6-10^7$~K, which varies as the gamma-ray luminosity changes, the sharp temperature spike in Fig.~1.

After the main bulk of gamma-rays passes through 
a given fluid element, the partial recombination occurs
producing mostly X-ray and soft X-ray photons. In a few minutes the
temperature drops down to $8-10\times 10^5$~K, and stays approximately at this level
until the gamma-ray front `causal discontiuity' crosses the shell 
and the thermal emission begins to escape through the outer boundary (i.e. the temperature radial profile becomes rather flat behind the heating front).

The momentum transferred by gamma-rays to matter pushes it outwards with velocities of about 3000~km~s$^{-1}$ 
The outward thermal emission flux  
is accumulated near the gamma-ray front, while the inward flux 
maintains thermal 
balance in the underlying zones, 
somewhat decelerating their motion (the innermost zones 
may be even pushed back by this flux towards the centre of explosion), 
and  then escapes through the inner boundary carrying the 
thermal energy out of the shell.

\begin{figure}
 \includegraphics[width=84mm]{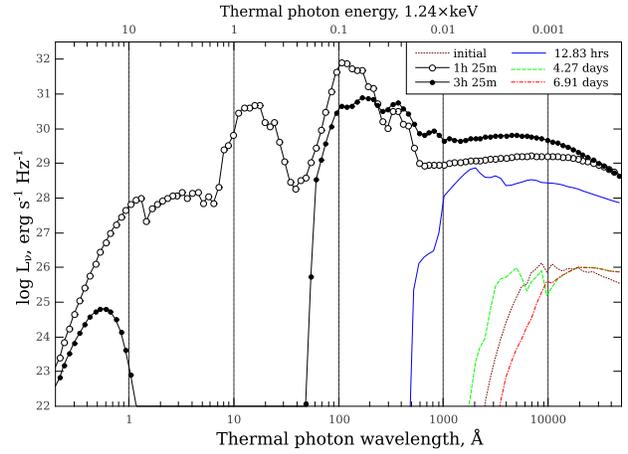}
\caption{Spectral evolution of the thermal luminosity 
from the GRB-heated circumburst shell. The curves 
correspond to different stages (peaks, breaks etc., compare the given times with the light curves in Fig.~2) after the forward gamma-ray escape,
i.e. after the GRB-trigger for an outer observer, except for the `initial' one which shows the unilluminated shell spectrum. }
\end{figure}

{\bf GRB ejecta effect}. After the jet kinetic energy has been injected (by means of the `thermal bomb', see the previous section), 
the medium gains an outward momentum and heat, and
a radiative shock is formed, with the temperature initially rising up to 
$1.5-3\times10^6$~K (see the temperature bump of $10^6$~K at $R_{[10^{14}]}\approx100.1$ in Fig.~1, corresponding to the time around 600~s after the `thermal bomb' had been 
activated).
The dynamics of matter is essentially controlled by the radiation flux 
since the radiation pressure is several orders of magnitude as 
high as that of the gas even at the shock front. Thus,  
thermal radiation from the shocked matter pushes forward external 
zones before the shock reaches them, 
and the inward radiation flux (summed with that coming from 
overlying radiatively heated zones) decelerates underlying shell zones. 
Fig.~1 illustrates how the velocity jump is smoothed out and a cooling front moves from the transparent inner boundary into the shocked shell.

As a result,  the shock is smeared out and its strength is damped, so 
the outer zones  
coast with a velocity of $1.1\times10^{4}$~km~s$^{-1}$ (in this model no external matter
to the shell is assumed). 

 \begin{figure}
 \includegraphics[width=84mm]{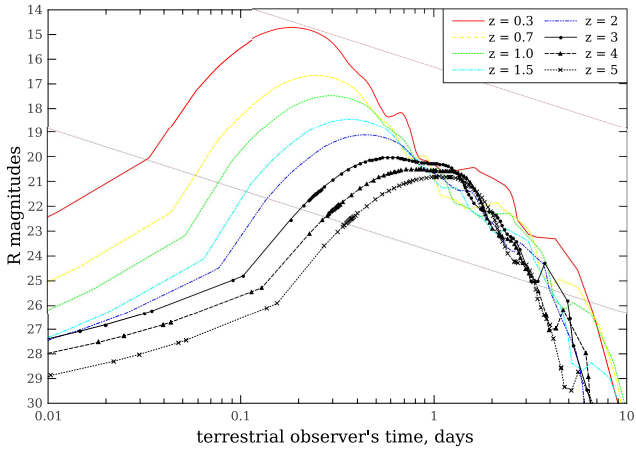}
 \caption{Calculated \textit{R}-fluxes of thermal 
emission seen from different redshifts. 
The straight dotted lines show the characteristic range 
of the observed \textit{R}-band afterglows of GRBs.}
\end{figure} 
 
{\bf Geometric curvature effect}. 
While the local radiation field intensity changes on time-scales of ionization ($\sim10^{-3}$~s), heating (several seconds) and cooling (minutes),
the observed light curve will also be subjected to light travel effects: the shell light crossing time and the geometric 
curvature delay, which in our case are $\sim1700$~s and $\sim5000$~s, respectively.

The shock emission becomes visible on the shell crossing time-scale (see optical curves in Fig.~2), while the bolometric luminosity maximum is achieved after
about the curvature delay time, when the whole hot shell area is observed. After that, a characteristic time-scale of one or another light curve variability 
 will reflect the cooling of matter in the shell (see e.g. a sharp break in the 2-30 keV curve in Fig.~2 indicating that the matter cools down to sub-keV temperature very quickly,
 during several seconds after the gamma-rays have escaped the shell and no longer supplied power to it).

Thus, due to rapid gamma-ray heating and radiative cooling of the shell,
a plateau with a steep decay emerges in the X-ray band. Thus, one should expect a more pronounced early and hard X-ray emission if the gamma-ray heating dominates. 
In the soft X-ray band a shallow (in logarithmic scale) rise  
and shallow but stepening decay will be observed, which 
can be interpreted in terms of superposition of 
the gamma-ray and the shock wave heating. 
The optical light curve is mainly shaped 
by the shock heating and thermal emission transport. This radiation should be more considerable if the most of the 
GRB energy comes in the jet kinetic energy. In extremal case where most of the kinetic energy is thermalized in the shell,
late-time bumps on the afterglow light curve may emerge, resembling a `quasi-supernova' as discussed below.

As can be seen in Fig.~4,  
the calculated R-fluxes of the optical thermal emission from sources at different redshifts
lie within the region occupied by the observed R-afterglows of GRBs. 
The colour of the synchrotron optical spectrum is red, 
therefore the flux in the rest-frame 
\textit{R}~band of the source is higher  
than that in \textit{R}~band of the terrestrial observer. 
As the calculated thermal emission optical spectrum is bluer than the synchrotron one 
(see Fig.~3), 
for a terrestrial observer the thermal emission increases 
on top of the `red' synchrotron radiation, which facilitates its detection.

The corresponding colour excess variations 
$\Delta(V-R)=(V-R)_{\rmn{tot}}-(V-R)_{\rmn{PLB}}$ are shown in Fig.~5 
for different strengths and spectral slopes of the power-law background 
(denoted as PLB, $F_\nu\sim\nu^{-\beta}$) for  
the source at moderate ($z=1$) and high ($z=6.29$) redshifts. 
The PLB spectral power indices are taken to be $\beta=0.7$ 
with $(V-R)_{\rmn{PLB}}=0.32$~mag and $\beta=1.3$ 
with $(V-R)_{\rmn{PLB}}=0.43$~mag. 
The brightest and the dimmest afterglow correspond to the upper and lower 
boundaries of
the observed power law afterglow range described above.
 
As can be seen from Fig.~5, during the first hours 
the afterglow appears to be `bluer' due to the contribution from thermal emission 
(although the total $V-R$ colour is still positive), 
and its colour 
remains constant during the light curve bump. After the spectral 
maximum passes through the 
(blueshifted) \textit{R} band, a noticeable reddening occurs. 
Apparently, the colour effect can be reliably observed only in afterglows with weak PLB.

\begin{figure}
 \includegraphics[width=84mm]{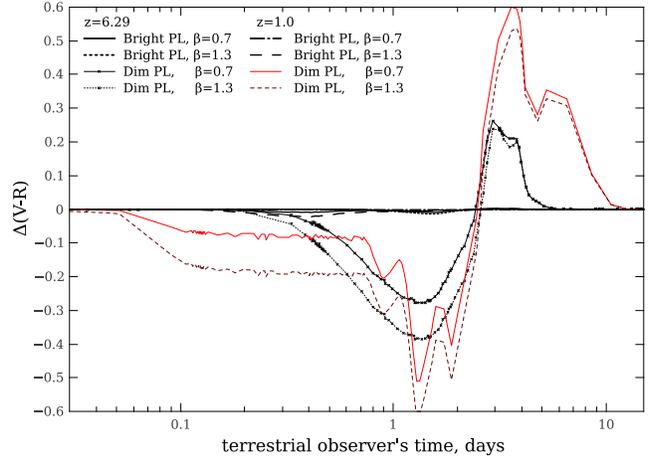}
 \caption{Colour evolution of the thermal bump for different 
underlying afterglow power-law forms and source redshifts. 
$\Delta(V-R)$ is the change in the $V-R$ colour 
relative to the PLB. The 'bright' and 'dim' 
PLB stand for the upper and lower boundaries of the observed $R$-afterglows in Fig.~4, 
respectively. The curves for bright afterglows are almost indisinguishable.}
\end{figure}

\begin{table*}
 \centering
 \begin{minipage}{120mm}
  \caption{Parameters of bumps of optical afterglow of GRBs 
obtained from data collected in \citet{BBG} and corrected for the Galactic extinction.}
  \begin{tabular}{@{}lllllll@{}}
  \hline
   GRB     & $z$   &  $E_{\gamma}$\footnote{Isotropic equivalent prompt emission energy in 1 keV -- 10 MeV range.}
                                      & $ R_{\rmn{bump}}$\footnote{\textit{R}-magnitude of the maximum}
                                                              & $t_{\rmn{peak}}$\footnote{Time since gamma-ray trigger (the peak luminosity).}  
                                                                       & $t_{\rmn{dur}}$\footnote{Duration of the bump}  
                                                                                   & $\beta$\footnote{Average spectral index in the \textit{R}-band, if available}   \\
           &       &  $10^{53}$ erg   & mag      & days      & days        &                \\
  \hline
  020124  & 3.198  & 1.6              & 18.36     & 0.47      & $\approx7$  & 0.56 \\
  021004  & 2.3351 & 0.1\footnote{Reduced to 15-150 keV range; only low-energy spectral fit was reported}
                                      & 16.2      & 0.08      & 0.5         & 0.67 \\
          &        &                  & 19.05     & 0.9       & 1.7         &      \\
  030328  & 1.52   & 3.3              & 19.4      & 0.28      & 1.2         & 0.36 \\
  030429X & 2.65   & 0.13\footnote{The luminosity was $5\times10^{53}$ erg s$^{-1}$}
                                      & 20.9      & 1.2       & $\approx4$  & 0.22 \\
  050904  & 6.29   & 6--32\footnote{Depending on different spectral peak energy estimations}
                                      & 20.5      & 0.32      & $\approx$7  & ...  \\
                                                                
   Model  & any    & 4\footnote{In 1 keV--30 MeV range}                      
                                      & Fig.~4 & Fig.~4 & Fig.~4   & Fig.~3,~5 \\

 \hline
\end{tabular}
\end{minipage}
\end{table*}

\section{Comparison with observations}

\textbf{Optical range}. 
Optical afterglows of several GRBs 
demonstrate clear deviations from the pure power-law decay 
(see e.g. discussion in Paper I). These 'bumps'  
correspond to additional proper luminosities of the order of
$10^{43}-10^{45}$~erg~s$^{-1}$, which is close to what is expected
from thermal emission of circumburst shells illuminated by prompt GRB emission.

Because of cosmological $K$-corrections the luminosity spectral density appears to be not a good 
estimator to compare observations and calculations due to 
cosmological $K$-corrections. 
Instead, we use the observed \textit{R}-band magnitudes, 
which can be derived from published data (corrected for the Galactic extinction),
and compare them with model $R$-magnitudes, which can be readily 
obtained as $(1+z)$ times shifted flux spectral density convolved 
with the \textit{R}-filter transmission function.

Table 1 lists properties of several GRBs which we use for comparison with our model. 
It shows that the observed bumps occur close to the time expected for 
thermal emission from the illuminated shell. 
For GRB 050904 no spectral index is available (it was not observed in \textit{R}-filter
because of its high redshift and 
the host galaxy Lyman-break absorption falling into $R$-band). 
The spectral slope and extrapolated \textit{R}-magnitudes 
(ignoring the host galaxy absorption) are estimated using 
available NIR colours (\citealt{TAC, PCM, HNR, BAD}, see also \citealt{KMK}).

 \begin{figure}
 \includegraphics[width=84mm]{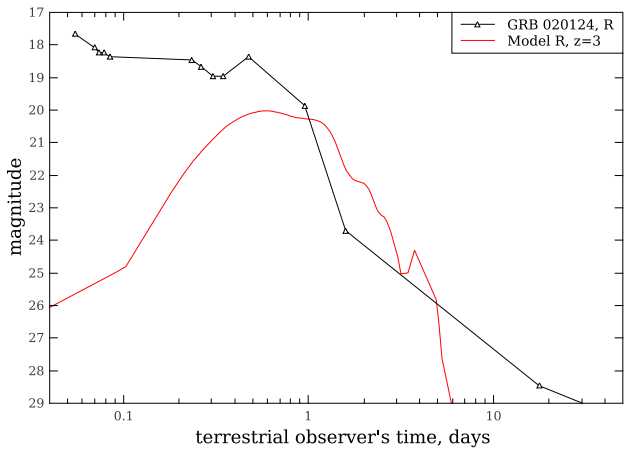}
 \caption{The GRB 020124 optical afterglow in the \textit{R} band (open circles) and the $z=3$ calculated thermal emission $R$-band stellar magnitudes (solid line).}
\end{figure}

Fig.~6 and 7 demonstrate  \textit{R}-light curves of GRBs 020124 and 050904 (synthesized)
with the calculated thermal emission superimposed. Clearly, the agreement is not perfect (we have not attempted to tune our model to fit any specific feature), 
but the similar time of emergence and amplitude of thermal emission from the gamma-ray illuminated circumburst shell in both cases seems to be suggestive.  

 \begin{figure}
 \includegraphics[width=84mm]{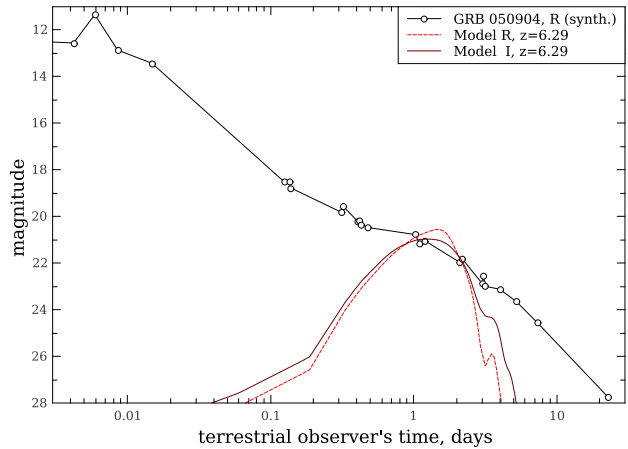}
 \caption{GRB 050904 optical afterglow extrapolated to \textit{R} band (open circles) and the $z=6.29$ calculated thermal emission stellar magnitudes in the $R$ (dashed line) and 
 $I$ (solid line) bands. }
\end{figure}

As can be seen from Fig.~6 and 7, the thermal emission does not correctly reproduce the onset of the irregularity. However, as \citet{NG07} show, such a smooth transition without 
a considerable rebrightening could be due to contribution of the shock synchrotron radiation when the shock encounters a sharp density contrast. The transition should become visible 
at times of several $T_0$ (the time when the shock strikes the dense region), i.e. on the time-scale of several $\delta t_{\gamma-\rmn{sh}}$ multiplied by a relevant $(1+z)$ factor.
The adopted $\delta t_{\gamma-\rmn{sh}}\sim10^2$~s, thus, gives appropriate times (since GRB trigger) of the irregularity onsets of $\sim0.1$~d. Later, when the thermal emission
gains strength and dominates, it forms long-term bump and rapid fading (which \citealt{NG07} argue to be not produced by the synchrotron emission only).

Note also that all considered GRBs have relatively low spectral index $\beta$, i.e. they 
tend to be systematically `bluer' than the typical power-law afterglows 
with mean $\beta\simeq0.7$\footnote{The spectral slopes have not been 
separately derived for the bumps; 
they have been taken from observational reports (see references in \citealt{BBG}), 
and hence are averaged over time.}. 
This also does not contradict their interpretation in terms of thermal emission
from circumburst shells.

\begin{figure}
 \includegraphics[width=84mm]{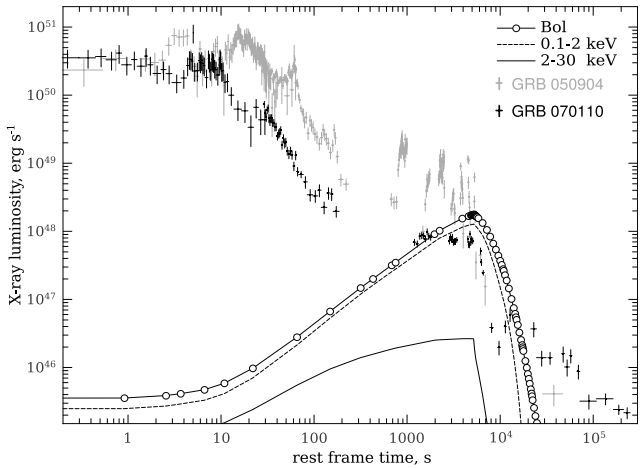}
 \caption{The GRB 050904 (grey crosses) and GRB 070110 (black crosses) X-ray luminosities obtained by \citet{TCB} from observed 0.3-10 keV fluxes, and model 
thermal emission light curves: bolometric (open circles), 0.1-2 keV (dashed line) and 2-30 keV (solid line). }
\end{figure}

\textbf{X-rays}. The calculated spectral density of thermal emission (see Fig.~2)
suggests that most of the shell thermal energy is 
carried away by X-rays. The properties of the emergent X-ray emission are 
directly sensitive to parameters of the shell 
and heating conditions, while the dependence of the optical emission
on the shell parameters is 
less straightforward to explain. 
The most notable feature of the thermal emission in X-rays 
is the appearance of a 'plateau' or slowly rising bumps 
mainly due to the high-latitude photon time delay from the shell's illuminated part.   

\citealt{TCB} present observed 0.3-10 keV afterglows of GRB 050904 and 070110 
converted into luminosity versus proper time, demonstrating a plateau exactly where the thermal bump is expected. In Fig.~8 we have placed both these light curves and 
our calculated bolometric, X-ray and soft X-ray ones. 

A thermal bump from the illuminated shell 
provides sufficient power to explain the feature observed in GRB 070110 afterglow,
but underestimates the amplitude of the X-ray plateau observed in GRB  050904 by 
an order of magnitude. However, GRB 050904 itself
was intrinsically brighter by an order of magnitude according to different estimates (see e.g. \citealt{KMK} and references therein).
The X-ray plateau in GRB 070110 had a flux of about 2~$\umu$Jy at 1~keV
corresponding to a luminosity spectral density of 
$L_\nu\approx5\times10^{29}$~erg~s$^{-1}$~Hz$^{-1}$ at 3.3~keV in the source frame, 
which is in good agreement with thermal emission model predictions (see Fig.~3). 
The calculated light curves also reproduce the abrupt end 
of both the plateaus occurring at the same time.

A certain discrepancy in the hardness ratio (1-10~keV to 0.2-1~keV) should be noted. 
\citet{TCB} reported that the observed hardness ratio of GRB 070110 was about one 
and increased up to $\approx1.5$ during the plateau phase, 
while the main part of our model thermal 
emission comes in soft X-rays (0.1--2 keV).

The deficit of hard photons in our model can be partially due to the one-temperature approximation. 
Indeed, the nearly instantaneous photoionization injects a lot of electrons with energies up to several ionization energies of ions they were knocked off, 
i.e. up to tens of keVs ($\sim10^8$~K), much higher than the ion temperature prior to the gamma-ray flashing. Compton scattering can additionally heat them up to 60~keV 
(the Compton temperature corresponding to the adopted gamma-ray spectrum and electron density, though this is only an approximate estimate, since the heating and cooling are 
non-stationary). Thus there should be more bremsstrahlung in X-ray and hard X-ray bands before they cool down to the equilibrium with ions ($10^5$~K).

Compton-scattered prompt emission photons should also noticeably contribute to hard X-rays and increase the hardness ratio. 
The prompt emission 1-60 keV photons scattered towards the observer can provide an X-ray luminosity of $L_{\leq60, \rmn{iso}}\approx1.25\times10^{49}$ erg s$^{-1}$ 
(assuming  the bolometric luminosity $3\times10^{53}$~erg/s). Also, some part of the inward coming thermal photons should be upscattered into hard X-rays.
The scattered emission will have a non-thermal spectrum harder than that shown in Fig.~3, and its light curve will show a rise on the curvature delay time-scale 
(i.e. thousands of seconds) followed by abrupt fading. This seems to be very similar to what is needed to produce the hard X-ray plateaus in the light curves of 
GRB 050904 and GRB 070110.

Thus, the observed hard X-ray irregularity, which can have a non-thermal origin, may indicate a shell-like circumstellar feature, 
which in turn may produce a thermal emission detectable as another bump in softer band at later time. 
Therefore we argue that the similarity between the observed afterglow bumps and the calculated 
thermal emission properties (in magnitudes, characteristic times, `bluer' spectral slopes, X-ray features) may be a strong indication of the presence of dense circumburst shells 
around some GRBs. This should be taken into account in studies of GRB afterglows.

\textbf{Effect of boundary conditions: the `quasi-supernova' effect.} 
When the inner boundary of the shell is rather opaque (or reflective, if there is a back scattering at the relativistic shock),
the radiation does not escape through it from the shell at all or 
diffuses slowly, while it can freely escape outwards. During the radiative heating phase (when the intermediate causal discontinuity exists)
the energy is trapped within hot parts of the shell (between the prompt emission front and the inner boundary) resulting in strong radiative forces. 
It dominates also after the energy injection  (`thermal bomb') 
due to the impact of the ejecta. During the first hours all motions of the hot medium are controlled by 
thermal radiation fluxes directed outwards. 

The main differences with the model considered above,
in which the inner boundary was transparent, are as follows. 
The 2--30 keV luminosity and its characteristic time-scale have not changed significantly,
but in soft X-rays and \textit{UBVRI} bands a bright flash appears produced by the shock impact, which becomes nearly ten times more powerful 
than the radiation produced in the case with the transparent inner boundary. 

This is followed by a several-day-long fading with the slope 
similar to the ordinary power-law afterglow. 
During this interval, there is no or only weak 
optical colour evolution, therefore the sum of 
thermal and non-thermal emission will behave as an ordinary afterglow 
(possibly slightly bluer than the usual one). 

 \begin{figure}
 \includegraphics[width=84mm]{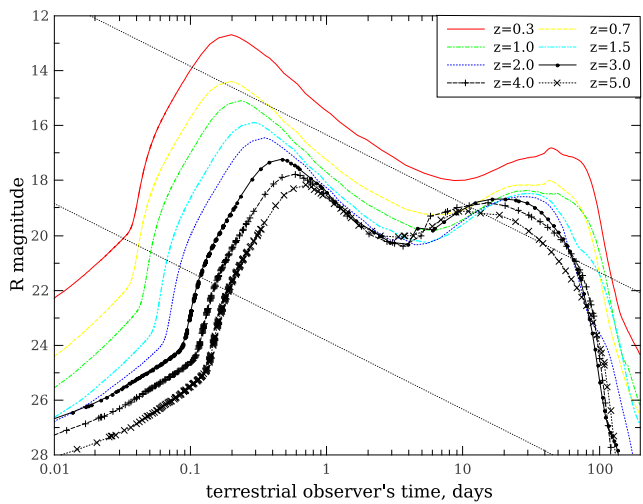}
 \caption{Calculated \textit{R}-fluxes of the `quasi-supernova' thermal emission from different redshifts. The straight dotted lines show the observed \textit{R}-band afterglow range.}
\end{figure}

The shock and the radiation pressure accelerate the shell up to $5-6.5\times10^4$~km~s$^{-1}$, so several days after the explosion, a bright supernova-like bump emerges, 
with colour changing from blue to red in a way similar to what is observed in a type IIn supernova originating in a dense circumstellar wind \citep{Fil97}. 
Model optical bumps observed from different redshifts are shown in Fig.~9.

Interestingly, similar double-bumped light curves were observed by 
\textit{Swift}-UVOT in GRB 060218 (SN2006aj) afterglow by \citet{Campana06}. 
A soft X-ray thermal component slowly rising for 3000 s and then fading abrubtly was also detected (see Fig.~2 in their paper). 
Rejecting a criticism cast on the feasibility of the shock breakout to produce a bright enough thermal emission, 
\citet*{Wax07} suggested the presence of an additional circumstellar structure (envelope) in the dense wind. 

Another evidence for a dense GRB environment was discussed by \citep{GSW}, who noted that
the SN2001ke assigned to the GRB 011121 was of type IIn rather than Ib/c.

Therefore, the radiation-driven `quasi-supernovae' might be responsible for late-time bumps in optical afterglows of GRBs, which are usually classified as supernova-bumps.
However in the present calculations, it appears about 4-5 magnitudes brighter than required to explain these bumps, and more detailed treatment of the inner boundary condition 
as well as non-spherical symmetric radiation transport is needed.

Either a strong opasity contrast or reflection is required to produce high radiation flux gradients driving the `quasi-supernova'.  
For example, high radiation flux gradients can appear due to high density contrasts in regions of the explosion, implosion, shocks
or cumulative jet developement, where the radiation or the ejecta propagate in a conical channel 
encircled by a cold dense medium, or there are clumps of matter on the way, when the jet protrudes through the progenitor star 
 or its extended atmosphere, when the blown-away material expands
into the stellar wind, or when the GRB radiation or ejecta perturb a filament of the circumburst matter, etc. Additionally, as mentioned above, the relativistic 
aberration of the emission scattered within the ejecta can effectively act like reflection.
In such cases supernova-like features can be observed in the afterglow, without supernova produced by the exploding star itself.  

It is tempting to apply this model to the explanation of the GRB-SN connection. 
While the GRB generating scenario is frequently referred to as a collapsar or a `failed supernova' \citep{W93}, a fairly `successful' supernova is required to produce observed 
late-time afterglow bumps.
If both phenomena originate in the progenitor star interiors, 
then the same GRB central engine must be able to drive both the narrow ultrarelativistic outflow with a low baryonic load, and the quasi-spherical explosion 
with a kinetic energy of the same order. Moreover, the GRB supernovae are found to be systematically intrinsically brighter than their regular mates 
(see e.g. section 4 in \citealt{Cano11}). 
However, this difficulty might be overcome if the central engine launches the jet only, and the source of the observed supernova `bump' lies outside the central star. 
In this case, the GRB supernova would be induced by the deposition of a portion of the GRB kinetic and/or radiation energy into the progenitor star exteriors 
or its environment.

Clearly, the problem of the GRB jet/shell interaction and the emerging radiation is multidimensional and relativistic, and so far no 
self-consistent solution has been obtained. 
Researchers working in the field either simplify (or neglect at all) the radiation transport, 
deriving the radiation properties from the hydrodynamics (e.g. \citealt{Lazzati09,Nagakura11}),
or limit themselves only by a simplified hydrodynamics (like the Blandford-McKee self-similar solution, see \citealt{TB03, Tolstov10}). 

We emphasize that 
thermal emission from circumburst medium 
can significantly contribute to the radiation of GRB afterglows.
The produced thermal radiation is strongly coupled  with properties of   
non-thermal emission of GRBs. Therefore, the development of 
sophisticated 
multifrequency multidimensional relativistic radiation hydrocodes to treat interaction of radiation with matter is
strongly desirable  to improve our understanding of the GRB physics.

\section{Conclusions}

In the present paper we have used the modified radiation hydrocode {\small STELLA} to calculate thermal radiation effects from the interaction of
the prompt GRB emission and GRB jet with the dense shells of matter which can surround massive GRB progenitors. Our model calculations revealed that  
thermal radiation of the heated circumburst structures 
can be visible in the X-ray and optical GRB afterglows in the form of plateaus and deviations from the pure power-law time decay. In particular, the calculated thermal radiatiton from 
a dense shell with the total mass of several $M_\odot$ located at a distance of $10^{16}$~cm from the GRB centre, which is  
illuminated by a 17-s long GRB with the total energy $4.5\times 10^{53}$~ergs, can reproduce both the time, shape and magnitude of 
real X-ray plateaus and irregularities of optical afterglows in several GRBs. It is found that the coupling between the radiation and 
dynamics of matter in such shells can lead to an interesting phenomenon -- a 'quasi-supernova' effect (radiation-driven explosion of the shell).
This intriguing possibility can explain the GRB-SN connection without requiring in some cases the quasi-spherical explosion of the progenitor star itself, 
and is worth further studying by means of more realistic multi-dimensional relativistic radiation hydrocodes.

\section*{Acknowledgements}
This work is supported by RFBR grant 13-02-92119. DAB thanks the support through the Dynasty foundation Support Program for Graduate Students and Young Pre-Degree Scientists. 
The work of SIB is supported partially by Russian grants RFBR 10-02-00249-A, Sci.~School   5440.2012.2, 3205.2012.2, 
IZ73Z0-128180/1 of the Swiss National Science Foundation (SCOPES) and the Grant
of the Government of the Russian Federation (No 11.G34.31.0047). The work of KAP is partially supported through RFBR grant 10-02-00599a.

\bsp

\label{lastpage}

\end{document}